\documentclass[journal]{IEEEtran}
\ifCLASSINFOpdf
\else
\fi
\usepackage{ifpdf}
\ifpdf
\usepackage[pdftex]{graphicx}
\graphicspath{{figs/}{matlab/}} 
\usepackage[update]{epstopdf}
\else
\usepackage{graphicx}
\graphicspath{{figs/}{matlab/}} 
\fi
\usepackage{array}
\usepackage{amsmath,amsthm}
\usepackage{amsfonts}
\usepackage{amssymb}
\usepackage{amstext}
\usepackage{latexsym}
\usepackage{xcolor}
\usepackage{cite,url}
\usepackage{setspace}

\usepackage[justification=centering]{caption}
\usepackage{float}
\newtheorem{definition}{Definition}
\newtheorem{theorem}{Theorem}
\newtheorem{corollary}{Corollary}
\usepackage{mathrsfs}
\usepackage[font=footnotesize]{caption}
\allowdisplaybreaks[4]

\begin{document}
\title{Secure Coded Caching with Colluding Users}

\maketitle

\begin{abstract}
In a secure coded caching system, a central server balances the traffic flow between peak and off-peak periods by distributing some public data to the users' caches in advance. Meanwhile, these data are securely protected against the possible colluding users, who might share their cache. We model the system as a flow network and study its capacity region via a network information-theoretic approach. Due to the difficulty of characterizing the capacity region straightforwardly, our approach is two folded from the perspective of network information theory. On one hand, we identify the inner bound of capacity region by proposing a coded caching scheme to achieve a low load secure data transmission. On the other hand, we also establish outer outer bounds on the capacity region, which show that our proposed scheme is order optimal in general under specific circumstance.

\end{abstract}


\IEEEpeerreviewmaketitle

\section{Introduction}\label{section1}
\IEEEPARstart{C}{aching} is an effective strategy to balance the traffic flow in communication networks. The data can be allocated in the storage units, or called as the cache, of each node during the off-peak periods. It can reduce the required transmission rate during peak periods, when the communication resources is scarce.

Recent studies have proposed the coded caching in information-theoretic framework \cite{caching1,tian2018caching,codingStorage1, codingStorage2}. In a coded caching systems, there is a central server distributing the data to some users through a broadcast channel. The goal is to minimize the amount of the required delivery content with the limited capacity of cache memories in every user. Meanwhile, some original data can sometimes be valuable, such as a paid video. The server doesn't want anyone who didn't pay for it to obtain these data. Therefore, the privacy and security of the data in coded caching have seized increasing interests in \cite{fundamentalLimits,ravindrakumar2016fundamental,zewail2016fundamental}. 
%In \cite{fundamentalLimits}, they guarantee the security of the distributing data from the eavesdroppers in the broadcast channel. The study in \cite{ravindrakumar2016fundamental,zewail2016fundamental} designs the coded caching schemes that can prevent the users to obtain the data that he didn't request.

We focus on a secure coded caching system with a general model. In this system, there is one central server connected to $K$ users through an error-free link. The server has a database of $N$ files, each with size $F$ bits. Every user can store no more than $MF$ bits of file data in his local cache. Before the transmission, each user will request one file without a priori and send the index of his request to server. Then, server broadcasts no more than $RF$ bits of file data to satisfy users' request. In addition, there are two constraints to protect the security of the data. When any $l$ users collude and share all the data in their cache, they can't obtain any information of all the original file data. Meanwhile, they can't obtain any information of the original files that they didn't request after the transmission.

In this paper, we call the two parameters $M$ and $R$ as the memory load and the communication load respectively. They are corresponding to the information on the node and the flow on the edge in the network. We aim to explore the region of all the possible rate pair $(M,R)$, to find the capacity of the network composed by a secure coded caching system. However, it is difficult to identify the all pairs, while the security constraints make this problem even harder.

The studies in \cite{caching1,tian2018caching} proposed the novel coded caching schemes and showed that they can be rather beneficial than uncoded schemes in such topology.
%\shao{"some studies" is a strange expression. The whole sentense also has something wrong. "suffer a great loss" means something bad.} 
Similar works also extended to the decentralized caching in \cite{codingStorage1, codingStorage2,distributedComputing}. Conversely, some studies made efforts to obtain theoretical lower bound of the necessary rate. The result in \cite{caching1} put up the lower bound, relying on the cut-set bound. In \cite{computeBounds3,computeBounds4}, the computer-aided methods are used to confirm the tighter lower bound.
%\shao{cut-set bound is a techinic, so it should be "applied" or used rather than studied.} 
These research investigated the fundamental limits of caching and helped to characterize the \emph{memory load-communication load} tradeoff of an optimal code.
%\shao{what do you mean by push?} 
When some addition security constraints are considered,  
%\shao{when xxx is considered}
\cite{fundamentalLimits} introduce the secret keys to encrypt the distributing data, so that the eavesdropper in the broadcast channel would fail to obtain the original files. The secret sharing schemes have been used to prevent every user to obtain the data that he didn't request in \cite{ravindrakumar2016fundamental} and \cite{zewail2016fundamental}. However, they didn't consider that a group of colluding users can easily decode the extra information with their cache. In this paper, we would further focus on the security with the presence of colluding users. If there might be up to $l$ of $K$ users colluding, the system should be able to limit any $l$ of the $K$ users in the entire process. If every user is rule-abiding and independent, the parameter $l$ would be $1$ and the system degenerates to that in \cite{ravindrakumar2016fundamental}, which indicates our model is more general. 

%Moreover, the security requirement in this system is using information-theoretic security raised by Shannon in \cite{shannonSecrecy}. The data will remain safe even if the eavesdroppers have unlimited power of computation.

%Our study of the capacity region is conducted from the perspective of network information theory. 

We have two main contributions in this paper. 
\begin{itemize}
    \item On one hand, we reveal the sufficient condition for an $(M,R)$ pair being achievable, which is called the inner bounds. We propose a coded caching scheme which can achieves this inner bound based on ideas from secret sharing\cite{cramer2012secure}.

    \item On the other hand, we establish the necessary condition of an $(M,R)$ being achievable, which is called the outer bounds. It also shows our inner bound is order optimal in general. 
\end{itemize} 
The inner and outer bounds together illustrate the capcity region.

This paper is organized as the following. In Section \ref{section2}, we propose a network information theory framework of the secure coded caching system. In Section \ref{section3}, we raise the inner bounds and outer bounds as our main results. In Section \ref{section4}, we explicitly shows our coded caching scheme, which can achieve our inner bounds for any general system scale. In Section \ref{section5}, we establish outer bounds, showing the optimality of our inner bounds. In Section \ref{section6}, we give out conclusion on this paper.

%Due to space constraints, we skip the detail proof of Section \ref{section5} here. These details may be found in \cite{}.

\section{Problem Formulation} \label{section2}
In this section, we formally propose the network information-theoretic model of the secure coded caching system. The model is defined in Section \ref{2A} and the corresponding code to realize the system function is defined in Section \ref{2B}.
	
\subsection{System Description}\label{2A}
\begin{figure} 
		\centering
		\includegraphics[width=0.8\linewidth]{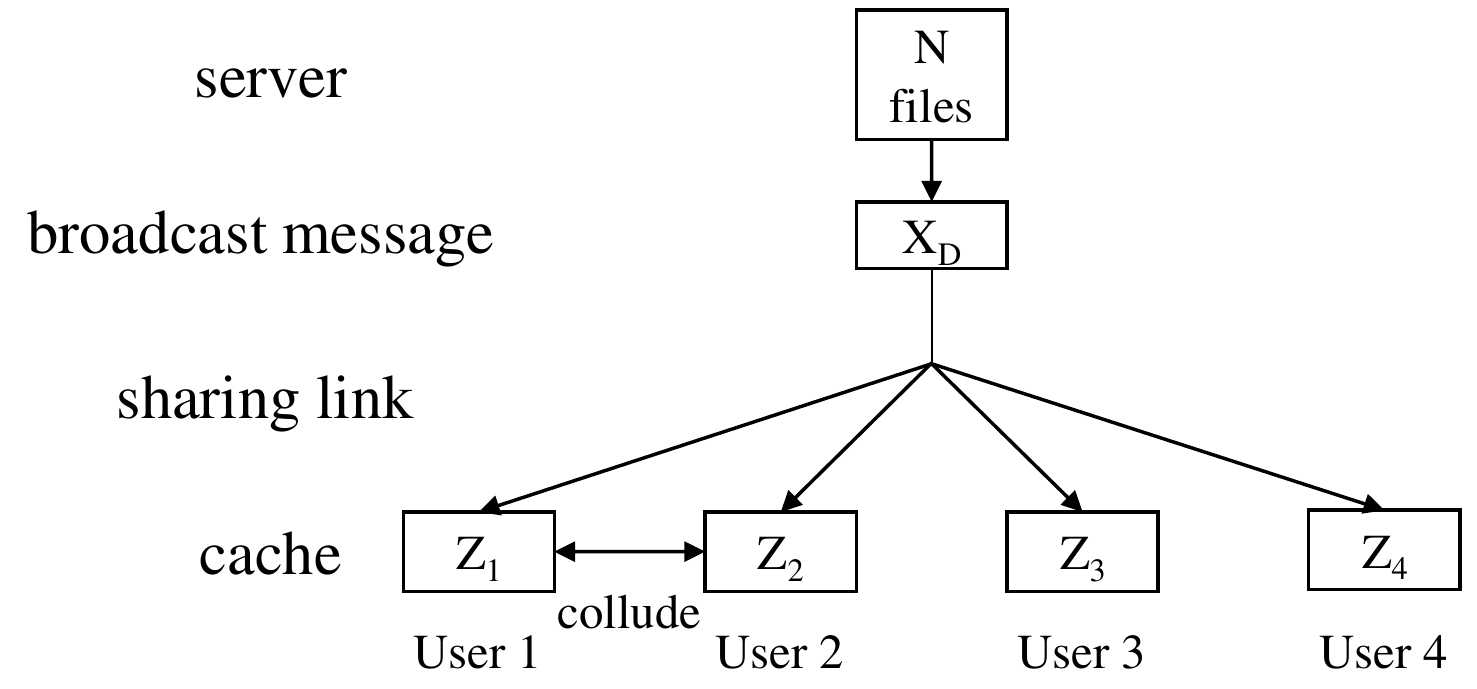} 
		\caption{A coded caching system with 4 users, where user 1 and user 2 collude.}
		\label{model_example}
		\vspace{-0.4cm}
\end{figure}

We consider a coded caching network with one central server that communicates with $K$ different users through an error-freely broadcast channel. There are $N$ distinct files in the database of the server. Besides, up to $l$ of these $K$ users might collude and share their known information with each other. As an example, Fig. \ref{model_example} shows such a coded caching system with 4 users, where any 2 users of them may collude in the communication.
	
Without losing generality, we regard the information of the $N$ files as some i.i.d random variables, denoted as $({W_n})_{n = 1}^N$. Each $W_n$ is uniformly distributed over
\begin{align*}
	[2^F]\buildrel \Delta \over = \{ 1,2,\dots,{2^F}\},
\end{align*}
where $F\in \mathbb{N}$. Each $W_n$ is large as $F$ bits. For the secrecy of the file data, we also introduce an encryption key $Y$ to avoid the data leakage between users and an encryption key $E$ to protect the data sharing. The encryption key $Y$ is a random variable to encrypt the $W_n$ and uniformly distributed over $\mathcal{Y}$. The encryption key $E$ is a random variable shared by users and uniformly distributed over $\mathcal{E}$. 
	
On the other hand, we regard the cache memory of the $k$th user $Z_k$ as a random variable as well, where $Z_k$ is uniformly distributed over $[2^{\lfloor M\!F\rfloor}]$. Each $Z_k$ is a collection of some different original files and random secret keys. We call $M$ as the memory load of each user, given the fact that $M$ indicates the total amount of data information possessed by each user.
	
Prior to data sharing, every user could request one of the $N$ files, and broadcast the index number of the requested file to the server. We use $d_{k}$ to denote the index number of the $k$th user's requested file. We can further define the demand vector $D=(d_{1},d_{2},\dots,d_{K})$. Under a given demand vector $D$, the server can functionally generate a corresponding broadcast message $X_D$, which is uniformly distributed over $[2^{\lfloor R\!F\rfloor}]$. After receiving the broadcast message, the $k$th user can decode $W_{d_k}$. Meanwhile, any $l$ of these $K$ users are possible to collude. But they can obtain no information of the other files which they didn't request. We call $R$ as the communication load of this system, since it indicates the number of bits transmitted by server under the worst case of task demand combination. 

As mentioned previously, our goal is to characterize the region of all the possible rate pair $(M,R)$. The definition of a rate pair being possible, or equivalently called as achievable, based on our information-theoretic model is introduced in the next subsection.
	
\subsection{Formal Statement}\label{2B}
In this part, we provide the formal definition of a secure caching code. An $(N,K,F,|\mathcal{Y}|,|\mathcal{E}|,M,R)$ secure caching code is consisted of:
\begin{itemize}
		\item for each $k\in [K]$, a memory caching functions 
		\begin{align*}
		\phi_{k}:[2^F]^N\times\mathcal{Y}\times\mathcal{E}\to[2^{\lfloor MF\rfloor}],
		\end{align*} 
		so that $Z_k$, the file data and encryption key cached by the $k$th user, is determined as
		\begin{align}
		Z_{k}=\phi_{k}(W_1,W_2,\dots,W_N,Y,E)\label{eq1};
		\end{align}
		
		\item for each $D\in [N]^K$, a message encoding function 
		\begin{align*}
		\psi_{D}:[2^F]^N\times\mathcal{Y}\times\mathcal{E}\to[2^{\lfloor RF\rfloor}],
		\end{align*}
		so that the broadcast message under given demand vector $D$ is determined as 
		\begin{align}
		X_{D}=\psi_{D}(W_1,W_2,\dots,W_N,Y,E)\label{eq2};
		\end{align}
		
		\item for $k\in [K]$ and $D\in [N]^{K}$, a decoding function
		\begin{align*}
		\mu_{k,D}:[2^{\lfloor RF\rfloor}]\times[2^{\lfloor MF\rfloor}]\to[2^F],
		\end{align*}
		so that for each user, he can decode his request file as
		\begin{align}
		W_{d_{k}}=\mu_{k,D}(Z_{k},X_{D}).\label{eq3}
		\end{align}
		
\end{itemize}
	
Furthermore, we define the set $\mathcal{L}=\{L\subset [K]:|L|=l \}$. We require the addition security constraints such that
%\begin{align}
\begin{gather}
	I(W_{[N]};\{Z_{k}:k\in L\})=0,\label{eq4}\\
	I(W_{[N]\setminus\{d_k:k\in L\}};X_{D},\{Z_{k}:k\in L\})=0,\label{eq5}
\end{gather}
%\end{align}
for any $L\in \mathcal{L}$ and $D\in [N]^{K}$. That is to say, for any $l$ users, colluding has no help on guessing the value of original file data. Moreover, no matter of their requests, colluding has no help on guessing the value of other unreqested file data after receving the the sharing message. The formal statement of rate pair being achievable is as follows:
	
\vspace{-0.15cm}
\begin{definition}
		The pair $(M,R)$ is achievable for an $(N,K)$ secure coded caching system if an $(N,K,F,|\mathcal{Y}|,|\mathcal{E}|,M,R)$ secure caching code can be found. We define the optimal achievable communication load under a certain memory load $M$ as $R^*(M)$ that
	\begin{align*}
		R^{*}(M) = \inf\{R:(M,R) \emph{ is achievable}\}.
	\end{align*}
\end{definition}
\vspace{-0.1cm}
	
Under any fixed system scale parameters, the achievable $(M,R)$ pairs consist a half open-half closed region. In order to find the fundamental limits, we propose inner bounds and outer bounds of the boundary on the closed part of the rate region. The rigorous expression of inner and outer bounds are in the next section.

\section{Main Results}\label{section3}
In this section, We present our main results. Theorem \ref{T1} presents the inner bounds of the \emph{memory load-communication load} tradeoff region of the secure coded caching system with general system scale parameters $(N,K)$.
	
\begin{figure}
		%\vspace{-0.8cm}  %调整图片与上文的垂直距离
		\setlength{\belowcaptionskip}{-0.5cm}   %调整图片标题与下文距离
		\centering
		\includegraphics[width=0.7\linewidth]{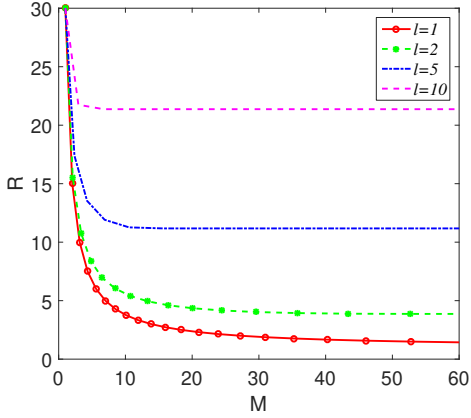}
		\caption{The communication load $R$ as a function of memory load $M$ in a secure coded caching system with $30$ users and $30$ tasks. The figure shows the performance of the coded scheme in Theorem \ref{T1} with $l=1,2,5,10$ colluding users.}
		\label{inner bound}
\end{figure}
	
\begin{theorem}\label{T1}
	In a secure coded caching system with the $N$ files and $K$ users with $l$ of whom might collude, a rate pair $(M,R)$ is achievable when
	\begin{align}
		&M=\frac{N\binom{K-1}{t-1}+\binom{K-1}{t}}{\binom{K-l}{t}},\label{T1-M}\\
		&R=\frac{\binom{K}{t+1}}{\binom{K-l}{t}}\label{T1-R}
		\end{align}
	for all $t\in \{0,1,\dots,\max(\left\lceil {\frac{{K + 1}}{l} - 2} \right\rceil,0)\}$, where $K,N\in \mathbb{N}$ and $l\in [K-1]$. Moreover, the lower convex envelope of these points is also achievable. 
\end{theorem}

The detailed coding scheme that achieves Theorem \ref{T1} is described and analyzed in the Section \ref{section4}. With the example of a secure coded caching system with $(N,K) = (30,30)$, we can show the achievable rate pairs by using coding theory in Fig. \ref{inner bound}. The several dash lines donate the bounds with different $l$. The common top extreme point $(1,30)$ can be achieved when each user caches one unique encryption key only. Using coding theory, we can achieve the region on and above these lines. We can define the function $R^C(M)$ to denote the corresponding achievable communication load rate with given $M$ according to Theorem \ref{T1}. 
		
Then we establish the outer bounds on the fundamental limits in Theorem \ref{T2}.
	
\begin{theorem}\label{T2}
In a secure coded caching system with the $N$ files and $K$ users, $l$ of whom might collude, for an achievable $(M,R)$ scheme that $M\geq 1$, we have $R \geq R_{s}^*(M)$
\begin{align}
		=\left\{
		\begin{array}{ll}
		\max \limits_{s\in\{l,l+1,\dots,\min(\lfloor N/2 \rfloor,K)\}}\displaystyle\frac{s\lfloor N/s\rfloor-l-(s-l)M}{\lfloor N/s \rfloor-1},\\
		{\qquad\qquad\qquad\qquad\qquad\qquad \qquad l< \min(\lfloor N/2 \rfloor,K)}\\
		\\
		\min(\lfloor N/2 \rfloor,K),
		{\qquad\qquad\quad \qquad l\geq \min(\lfloor N/2 \rfloor,K})\\
		\end{array}\right.	
		\label{RS}
\end{align}
where $K,N\in \mathbb{N}$ and $l\in [K-1]$.
\end{theorem}	
Theorem \ref{T2} gives out the necessary condition on the communication load. The function $R_{s}^*(M)$ denotes one lower bound of $R^*(M)$ which is concluded from many outer bounds. The detailed proof of Theorem \ref{T2} are presented in the Section \ref{section5}. With Theorem \ref{T2}, we can induce the optimality of our inner bound as following Corollary.

\begin{figure} 
	\centering
	\includegraphics[width=0.75\linewidth]{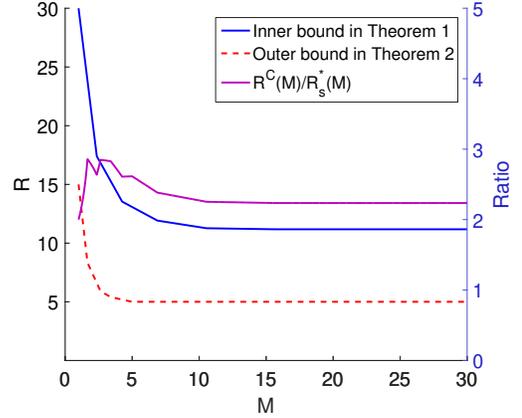} 
	\caption{The comparision between the inner bound in Theorem \ref{T1} and the outer bound in Theorem \ref{T2} in an $(N,K)=(30,30)$ system with $l=5$.}
	\label{Ratio Comparision}
	\vspace{-0.5cm}
\end{figure}
	
\begin{corollary}\label{C1}
		In a secure coded caching system with the $N$ files and $K$ users, $l$ of whom might collude,
		\begin{align}
		1\le\frac{R^{C}(M)}{R^{*}(M)} \le 12,
		\end{align}
		for every achievable rate pair $(M,R^C(M))$ in Theorem \ref{T1} when $t\geq\lfloor \frac{K+1}{10l} \rfloor$ and $\frac{K}{N}\leq 3$. 
\end{corollary}
Corollary \ref{C1} shows that the inner bound in Theorem \ref{T1} is order optimal for most regimes of interest. The detailed proof of Corollary \ref{C1} is in Section \ref{section5}. The Fig. \ref{Ratio Comparision} shows the comparison of the inner bound and outer bound of an $(N,K)=(30,30)$ system with $l=5$. The optimal fundamental limits lie in the gap between the red dashed line and the blue line, hence the ratio between $R^{C}(M)$ and $R^{*}(M)$ can be upper bounded by $R^C(M)/R_s^*(M)$. We compare the ratio of $R^C(M)/R_s^*(M)$ with the same $M$. We can see the ratio is always smaller than 3. It indicates the order optimality of our inner bound in the $(N,K)\!=\!(30,30)$ system with $l=5$.

\section{Caches Assignment and Communication Scheme}\label{section4}
In subsection \ref{4-A}, we put up the coding scheme and prove Theorem \ref{T1}. In subsection \ref{4-B}, a example would help to illustrate the proposed scheme.	

\subsection{The General Caching Scheme}	\label{4-A}
We aim at generalizing the caching strategy to prove the inner bound in Theorem 1, in terms of $K$, $N$, and parameters, $l$ and $t$. Consider a caching system with the $N$ files and $K$ users. We know there might be up to $l$ untrustworthy users, and $l\in [K-1]$. 
		
\subsubsection{File Precoding}\qquad
	
Firstly, we split each file into $P=\binom{K-l}{t}$ subfiles of equal size, where $t\in \{0,1,\dots,\max(\left\lceil {\frac{{K + 1}}{l} - 2} \right\rceil,0)\}$. These subfiles are distinct and independent, with size of $F/P$ bits each. We define $\{W_{n,p}:n\in[N], p\in [P]\}$ to be $NP$ independent random variables each uniformly distributed over $[2^{F/P}]$. They can denote the information of these subfiles. i.e., $W_n=(W_{n,1},W_{n,2},\dots,W_{n,P})$ for every $n \in [N]$.
	
Then, we would generate $NQ=N[\binom{K}{t}-\binom{K-l}{t}]$ independent random encryption keys $\{Y_{n,q}:n \in[N], q\in [Q]\}$. Each $Y_{n,q}$ is uniformly distributed over $[2^{F/P}]$ and independent of the files, with size of $F/P$ bits each. These keys would be used to encrypt the file blocks.
		
Finally, we define $NG=N(P+Q)$ independent random variables $\{\tilde{W}_{n,\mathcal{T}}:n \in[N], \mathcal{T} \subseteq [K] \text{ and }|\mathcal{T}|=t\}$ to denote the file blocks, each uniformly distributed over $[2^{F/P}]$. Meanwhile, we set $\tilde{W}_{n,\mathcal{T}_i}=\sum_{p=1}^{P}{W_{n,p}c_{i,p}+\sum_{q=1}^{Q}{Y_{n,q}c_{i,q+P}}}$, where $i$ denotes the index of $\mathcal{T}_i$ in the set $\{\mathcal{T} \subseteq [K]   , |\mathcal{T}|=t\}$ and each $c_{i,j}$ is the element in a $G \times G $ Cauchy matrix, which is constructed in Galois field. These blocks form a $(G,P,Q)$ secret sharing scheme based on the Cauchy Reed-Solomon codes \cite{plank2005optimizing}. We can induce some properties as following Corollary. This fact is proved in appendix.%\cite{}.
	
\begin{corollary}\label{C2}
	For each $n\in [N]$,
	\begin{itemize}
			\item [a)]
			All $G$ blocks of $W_{n}$ could recover the orginal file ${W}_{n}$.
			\begin{align}
			H(W_{n}|\{\tilde{W}_{n,\mathcal{T}_{1}},\tilde{W}_{n,\mathcal{T}_{2}},...,\tilde{W}_{n,\mathcal{T}_{G}}\})=0.\label{C2eq1}
			\end{align} 
			\item [b)] 
			Any $Q$ or less blocks of $W_{n}$ would not leak any information about the orginal files.    
			\begin{align}
			I(W_{[N]};\{\tilde{W}_{n,\mathcal{T}_{i_{1}}},\tilde{W}_{n,\mathcal{T}_{i_{2}}},...,\tilde{W}_{n,\mathcal{T}_{i_{Q}}}\})=0,\label{C2eq2}
			\end{align} 
			where $\{i_1,i_2,\dots,i_Q\}$ is one subset of $[G]$.
	\end{itemize}	
\end{corollary}

\subsubsection{Placement Strategy}\qquad
	
After precoding files, the server would generate some independent random sercet keys $\{E_{\mathcal{T}^+}:\mathcal{T}^+ \subseteq [K] \text{ and } |\mathcal{T}^+|=t+1\}$. These keys are uniformly distributed over $[2^{F/P}]$ and independent of the files, with size of $F/P$ bits. These keys would be used to encrypt the broadcast symbols later. 

For the prefetching, each user $k$ caches all the blocks $\tilde{W}_{n,\mathcal{T}}$ satisfying $k \in \mathcal{T}$. He need aslo caches all the unique keys $E_{\mathcal{T}^+}$ satisfying $k \in \mathcal{T}^+$. Thus, we can conclude the $Z_{k}$ from above
\begin{align}
	Z_{k}=\{   \tilde{W}_{n,\mathcal{T}}      :    n\in [N],    k\in\mathcal{T}\}\cup 
	\{E_{\mathcal{T}^+}:k\in \mathcal{T}^+ \}.\label{Zk}
\end{align}
	
\subsubsection{Transmission Strategy}\qquad
	
Before data sharing, every user broadcasts the index number of the requested file. For the received demand vector $D=(d_{1},d_{2},\dots,d_{K})$, server will transmit these symbols:$X_{D,\mathcal{T}^+}=E_{\mathcal{T}^+}\oplus\underset{u\in\mathcal{T}^+}{\bigoplus}\tilde{W}_{d_u,\mathcal{T}^+\setminus{u}}$, for each sets $\mathcal{T}^+ \subseteq [K] \text{ and } |\mathcal{T}^+|=t+1$. These symbols form the message 
\begin{align}
	X_D=\{X_{D,\mathcal{T}^+}:\mathcal{T}^+ \subseteq [K] \text{ and } |\mathcal{T}^+|=t+1\}.\label{XD}
\end{align}

\subsubsection{Proof of Achievability}\qquad

In the general scheme, we design $Z_{k}$ and $X_{D}$ as \eqref{Zk} and \eqref{XD}. To confirm the achievability of the proposed scheme, we would continue to prove every user can recover his requested file after receiving the transmission content and secrecy constraints \eqref{eq4} and \eqref{eq5} are satisfied.
	
Without loss of generality, let us see how the $k$th user get $W_{d_{k}}$ after receiving all broadcast message. From the cache, he could obtain some file blocks $\{\tilde{W}_{d_k,\mathcal{T}}:k\in\mathcal{T}\}$. For every $\mathcal{T}$ that has $k\notin\mathcal{T}$, he collects the transmission symbol $\{X_{D,\mathcal{T}^+}:\mathcal{T}^+=\{k\cup\mathcal{T}\}\}$. Meanwhile, we have $
X_{D,\mathcal{T}^+}\!=\!X_{D,k\cup\mathcal{T}}=\underset{u\in\{k\cup\mathcal{T}\}}{\bigoplus}\tilde{W}_{d_u,{k\cup\mathcal{T}}\setminus{u}}\oplus E_{{k\cup\mathcal{T}}}=[\underset{u\in\mathcal{T}}{\bigoplus}\tilde{W}_{d_u,{k\cup\mathcal{T}}\setminus{u}}\oplus E_{{k\cup\mathcal{T}}}]\oplus\tilde{W}_{d_k,\mathcal{T}}$.

For user $k$, it is clear that only $\tilde{W}_{d_k,\mathcal{T}}$ is useful, while the rest blocks and unique keys are interference signals. We can see that these interference signals were stored in the user $k$'s cache. Thus, he could recover the block $\tilde{W}_{d_k,\mathcal{T}}=X_{D,\mathcal{T}^+}\oplus[\underset{u\in\mathcal{T}}{\bigoplus}\tilde{W}_{d_u,{k\cup\mathcal{T}}\setminus{u}}\oplus E_{{k\cup\mathcal{T}}}]$, for every $\mathcal{T}$ with $k\notin\mathcal{T}$. Hence, He could obtain the rest blocks $\tilde{W}_{d_k,\mathcal{T}}$ with $k\notin\mathcal{T}$
After obtaining all blocks of ${W}_{d_k}$, user $k$ can recover the orginal file ${W}_{d_k}$ according to the Corollary \ref{C2}.
	
Then we consider the worst case that there are $l$ colluding users, who are willing to share the information with each others. They may try to get some extra file information by cooperating, from which might make them profit. Our target is to prevent them to get any file or information that they didn't request. We assume the $l$ colluding users are user $j_{1},j_{2},\dots,j_{l}$, where set $L=\{j_{1},j_{2},\dots,j_{l}\}$ is one subset of $[K]$. Before the delivery, the $l$ colluding users may share their cache. We denote the union set of their cache as $Z_{L}$, which contains two parts $Z_{L,W}$ and $Z_{L,E}$:
\begin{align*}
	Z_{L,W}=\{&\tilde{W}_{n,\mathcal{T}}:n\in[N],\mathcal{T}\cap L\ne \emptyset\}\\
	Z_{L,E}=\{&E_{\mathcal{T}^+}:\mathcal{T}^+\cap L\ne \emptyset\}
\end{align*}	
For each $n \in [N]$, the number of blocks $\tilde{W}_{n,\mathcal{T}}$ they can get is $[\binom{K}{t}-\binom{K-l}{t}]=Q$. Thus, they could never get any information about any files before they request the file according to the Corollary \ref{C2}. Equation \eqref{eq4} is satisfied.
	
After the delivery, the $l$ colluding users may share their cache again. We divide the message $X_D$ into two sets: $X_{1}=\{X_{D,\mathcal{T}^+}:\mathcal{T}^+\cap L \ne \emptyset\}$ and $X_{2}=\{X_{D,\mathcal{T}^+}:\mathcal{T}^+ \cap L=\emptyset\}$. It is obvious $X_{2}$ is no help for them to guess any information about files, because every symbol $X_{D,\mathcal{T}^+}\in X_{2}$ has been encrypted by the unique key $E_{\mathcal{T}^+}$ that is not in set $Z_{L,E}$. On the other hand, for every symbol $X_{D,\mathcal{T}^+}\in X_1$, we could find a number $j$ satisfied $j\in L$ and $j\in \mathcal{T}^+$. We have $X_{D,\mathcal{T}^+}=E_{\mathcal{T}^+}\oplus\underset{u\in\mathcal{T}^+}{\bigoplus}\tilde{W}_{d_u,\mathcal{T}^+\setminus{u}}=\tilde{W}_{d_j,\mathcal{T}^+\setminus{j}}\oplus E_{\mathcal{T}^+}\oplus\underset{u\in\mathcal{T}^+\setminus j}{\bigoplus}\tilde{W}_{d_u,\mathcal{T}^+\setminus{u}}$.
As we analyze before, for user $j$, only $\tilde{W}_{d_j,\mathcal{T}^+\setminus{j}}$ is useful. The rest blocks and unique keys are interference signals. We know these interference signals were stored in his cache. The $l$ colluding users could only obtain $\tilde{W}_{d_j,\mathcal{T}^+\setminus{j}}$, which user $j$ deserves. It means they fail to get one more blocks of unrequested files from $X_1$. Thus, for each $W_n$ they didn't request, the $l$ colluding users can only acknowledge $Q$ blocks. They could never get any information about these files according to the Corollary \ref{C2}. Equation \eqref{eq5} is satisfied.
	
\subsubsection{The Rate Pair of the Proposed Scheme}\qquad
	
Now we comfirm our scheme is achievable. We can continue to determine the rate pair $(M,R)$. From \eqref{Zk}, every user caches $N\binom{K-1}{t-1}$ blocks and $\binom{K-1}{t}$ unique keys. Thus the memory load of the user: $M=\frac{N\binom{K-1}{t-1}+\binom{K-1}{t}}{\binom{K-l}{t}}$. From \eqref{XD}, server transmits $\binom{K}{t+1}$ symbols $X_{D,\mathcal{T}^+}$. Hence, the  communication  load $R=\frac{\binom{K}{t+1}}{\binom{K-l}{t}}$. We can obtain the rate pair $(M,R)$ as defined in Theorem 1. Theorem \ref{T1} is proved.	
	
\subsection{Example}	\label{4-B}
$(N=4,K=4,l=2)$ Consider a coded caching system with $4$ files, $W_1,W_2,W_3,W_4$, and $4$ users. We know any of two could collude. 
%Here, we set $t=1$. 
Firstly, we divide every file into two partitions. i.e. $W_{i}=(W_{i,1},W_{i,2})$ for every $i \in \{1,2,3,4\}$. Then, we generate some encryption keys and unique keys: $Y_{1,1}$,$Y_{1,2}$,$Y_{2,1}$,$Y_{2,2}$,$Y_{3,1}$,$Y_{3,2}$,$Y_{4,1}$,$Y_{4,2}$, $E_{12}$,$E_{13}$,$E_{14}$,$E_{23}$,$E_{24}$,$E_{34}$. We use a Cauchy matrix to form the blocks as below:

\begin{footnotesize}
\[\begin{array}{l}
\left[ {\begin{array}{*{20}{c}}
	{{{\tilde W}_{i,1}}}\\
	{{{\tilde W}_{i,2}}}\\
	{{{\tilde W}_{i,3}}}\\
	{{{\tilde W}_{i,4}}}
	\end{array}} \right] = \left[ {\begin{array}{*{20}{c}}
	7&6&8&2\\
	6&8&2&12\\
	8&2&12&5\\
	2&12&5&10
	\end{array}} \right]\left[ {\begin{array}{*{20}{c}}
	{{W}_{i,1}}\\
	{{W}_{i,2}}\\
	{{Y_{i,1}}}\\
	{{Y_{i,2}}}
	\end{array}} \right],
\end{array}\]
\end{footnotesize}
where $i \in \{1,2,3,4\}$. The cache at the every user are:
\begin{align*}
	Z_{1}=\{{\tilde W}_{1,1},{\tilde W}_{2,1},{\tilde W}_{3,1},{\tilde W}_{4,1},E_{12},E_{13},E_{14}\},\\
	Z_{2}=\{{\tilde W}_{1,2},{\tilde W}_{2,2},{\tilde W}_{3,2},{\tilde W}_{4,2},E_{12},E_{23},E_{24}\},\\
	Z_{3}=\{{\tilde W}_{1,3},\tilde W_{2,3},\tilde W_{3,3},\tilde W_{4,3},E_{13},E_{23},E_{34}\},\\
	Z_{4}=\{{\tilde W}_{1,4},\tilde W_{2,4},\tilde W_{3,4},\tilde W_{4,4},E_{14},E_{24},E_{34}\}.
\end{align*}

Here we assume the demand vector $D=(1,2,3,4)$, server will transmit 6 symbols:
\begin{align*}
	X_{D}=\{\tilde W_{2,1}\oplus \tilde W_{1,2}\oplus E_{12},
			\tilde W_{3,1}\oplus \tilde W_{1,3}\oplus E_{13},&\\
	    	\tilde W_{4,1}\oplus \tilde W_{1,4}\oplus E_{14},
		    \tilde W_{3,2}\oplus \tilde W_{2,3}\oplus E_{23},&\\
	    	\tilde W_{4,2}\oplus \tilde W_{2,4}\oplus E_{24},
	    	\tilde W_{4,3}\oplus \tilde W_{3,4}\oplus E_{34}.&\}
\end{align*}

Every user could decode all the file blocks of their requested file and recover the file. For user 1, he wants to recover $W_1$.
He can directly obtain the block ${\tilde W}_{1,1}$ from cache $Z_1$. The rest blocks of his request file $W_1$ could be decoded from $X_D$:
\begin{align*}
	{\tilde W}_{1,2}=[\tilde W_{2,1}\oplus \tilde W_{1,2}\oplus E_{12}]\oplus \tilde W_{2,1}\oplus E_{12},\\
	{\tilde W}_{1,3}=[\tilde W_{3,1}\oplus \tilde W_{1,3}\oplus E_{13}]\oplus \tilde W_{3,1}\oplus E_{13},\\
	{\tilde W}_{1,4}=[\tilde W_{4,1}\oplus \tilde W_{1,4}\oplus E_{14}]\oplus \tilde W_{4,1}\oplus E_{14}.
\end{align*}	
Now he have get ${\tilde W}_{1,1},{\tilde W}_{1,2},{\tilde W}_{1,3},{\tilde W}_{1,4}$. He could figure out the subfiles $W_{1,1},W_{1,2}$ and recover $W_1$. Similarly, other users could obtain their request file.
	
If any of two collude, they could not figure out any information from the unquested files. Supposing user 1 and user 2 share their information. Before delivery, they can't eliminate the interference of encryption key $Y_{n,p}$ with only 2 blocks of each file from cache. They can't get any information about any files. After delivery, they can't decode $\tilde W_{4,3}\oplus \tilde W_{3,4}\oplus E_{34}$ due to the interference of $E_{34}$. Thus, they fail to obtain one more block of unrequest files and can't get any extra information.
	
\section{Proof of Outer Bounds}\label{section5}
We give the derivation of the outer bound in Theorem \ref{T2}, similar to \cite{fundamentalLimits}. We also compare the achievable rate $R^C(M)$ of the proposed scheme with the optimal achievable communication load rate $R^*(M)$. The result shows that the ratio between them is is within a constant factor of the optimal for most regimes of interest. The proof can be found in appendix.%\cite{}.

\section{Conclusion}\label{section6}
In this paper, we study the fundamental limits of the \emph{memory load-communication load} tradeoff region for secure coded caching system with colluding users. On one hand, we propose a coded caching scheme using coding theory, which can have a lower communication load under a same memory load, comparing with no coding theory used scheme. On the other hand, we also established outer bounds for the system. The outer bounds show that our proposed scheme is order optimal in general. Although larger rate needed to achieve the goals compared with the optimal rate, we could control the costs within the tolerable range for most regimes.

\bibliographystyle{./bibliography/IEEEtran}
%\bibliography{./bibliography/IEEEabrv,./bibliography/IEEEexample}
\bibliography{./reference}

\newpage
\appendices
\section{Proof of Theorem \ref{T2}}\label{BA}
In this part, we prove the outer bound in Theorem \ref{T2}, similar to \cite{fundamentalLimits}. Consider an achievable $(M,R)$ scheme with the $N$ files and $K$ users. We would calculate the lower bound of the optimal achievable communication load rate $R^*(M)$. Firstly, we consider the case of $l< \min(\lfloor {N/2} \rfloor,K) $. Let $s$ be a integer that $s\in\{l,l+1,\dots,\min(\lfloor {N/2} \rfloor,K)\}$. We assume there are $\lfloor N/s \rfloor⌋$ possible demand vectors: $D_1, D_2,\dots, D_{\lfloor N/s\rfloor}$. At the first request instance, the first $s$ users request the files $W_1,W_2,\dots,W_s$. At the second request instance, the first $s$ users request the files $W_{s+1},W_{s+2},\dots,W_{2s}$. Proceed in the same law. At the $p$th request instance, the first $s$ users request the results $W_{(p-1)s+1},W_{(p-2)s+2},\dots,W_{ps}$. i.e., the first $s$ elements in $D_p$ is $(p-1)s\!+\!1,(p-1)s\!+\!2,\dots,ps$, where $p \in \{1,2,\dots,\lfloor N/s \rfloor⌋\}.$ We keep to set some expression as below:
\begin{align*}
	\tilde X &= \{ X_{D_1},X_{D_2},...,X_{D_{\left\lfloor {N/s} \right\rfloor }}\}, \\
	{{\tilde X}_{\backslash p}} &= \{ X_{D_1},...,X_{D_{p-1}},X_{D_{p+1}},....,X_{D_{\left\lfloor {N/s} \right\rfloor }}\}, \\
	\tilde Z &= \{ {Z_1},{Z_2},...,{Z_s}\}, \\
	\tilde W &= \{ {W_1},...,W_{(p-1)s},W_{(p-1)s+l+1}...,{W_{\left\lfloor {N/s} \right\rfloor s}}\}.
\end{align*}
Here $\tilde X $ denotes the set of all the broadcast message at the all $\lfloor N/s \rfloor$ request instances. ${{\tilde X}_{\backslash p}} $ denotes the set of all the broadcast message at the all but the $p$th request instances. Because $s\le \lfloor {N/2} \rfloor$, $\lfloor {N/s} \rfloor\ge 2$. ${{\tilde X}_{\backslash p}} $ is always non empty. $\tilde Z$ denotes the union set of the first $s$ users' cache. $\tilde W$ denotes the union set of the requested files at the all request instances except $W_{(p-1)s+1},W_{(p-1)s+2},\dots,W_{(p-1)s+l}$, which are the requested files of the first $l$ users at the $p$th request instance. 
	
Meanwhile, we can generalize some constraints based on the definition of the system. We know all the requested files in $\tilde W$ could be recovered if given $\tilde X$ and $\tilde Z$, we have
\begin{align}
	H(\tilde W|\tilde X,\tilde Z) = 0,\label{eq19}.
\end{align}

Moreover, the secrecy constraint should be satisfied. We should prevent the first $l$ users to obtain the information of the unrequested files at the $p$th request instance. Thus, according to \eqref{eq5}, we have
\begin{align}
	I(\tilde W;{X_p},Z_1,Z_2,\dots,Z_l) = 0,\label{eq20}
\end{align}
where $p \in \{ 1,2,...,\left\lfloor {N/s} \right\rfloor \} $. Therefore we now give derivation below:
\begin{align}
	(\lfloor {N/s}\rfloor s-l)F=& H(\tilde W) \notag\\
	=& I(\tilde W;\tilde X,\tilde Z) + H(\tilde W|\tilde X,\tilde Z)\notag\\
	\overset{(a)}{=}& I(\tilde W;\tilde X,\tilde Z) \notag\\
	=& I(\tilde W;{X_p},Z_1,\dots,Z_l) \notag\\
	&+ I(\tilde W;{{\tilde X}_{\backslash p}},Z_{l+1},\dots,Z_s|{X_p},Z_1,\dots,Z_l)\notag\\
	\overset{(b)}{=}& I(\tilde W;{{\tilde X}_{\backslash p}},Z_{l+1},\dots,Z_s|{X_p},Z_1,\dots,Z_l)\notag\\
	\le& H({{\tilde X}_{\backslash p}},Z_{l+1},\dots,Z_s)\notag\\
	\le& \sum\limits_{j=1,j\not=p}^{\lfloor {N/s}\rfloor}H(X_{D_j})+\sum\limits_{i=l+1}^{s}H(Z_{i})\notag\\
	=&(\left\lfloor {N/s} \right\rfloor  - 1)RF + (s-l)MF.\label{eq21}
\end{align}
Here $(a),(b)$ are due to the constraints in \eqref{eq19},\eqref{eq20}. Therefore, from \eqref{eq21}, we can get a series of outer bounds,
\begin{align}
	(\left\lfloor {N/s} \right\rfloor  - 1)R + (s-l)M\ge \lfloor {N/s}\rfloor s-l ,\label{eq22}
\end{align}
where $s\in\{l+1,l+2,\dots,\min(\lfloor {N/2} \rfloor,K)\}$. The lower bound of the optimal achievable communication load rate ${R^*}(M)$ could be concluded from these outer bounds
\begin{align}
	R^*(M)&\ge R_s^*(M)\notag\\
	&=\max \limits_{s\in\{l,l+1,\dots,\min(\lfloor N/2 \rfloor,K)\}}\displaystyle\frac{s\lfloor N/s\rfloor-l-(s-l)M}{\lfloor N/s \rfloor-1}.\label{eq26}
\end{align}
	
Then, we consider the case of $l\geq \min(\lfloor {N/2} \rfloor,K)$. Let $s$ be a integer that $s\in\{1,2,\dots,\min(\lfloor {N/2} \rfloor,K)\}$. We still assume there are $\lfloor N/s \rfloor⌋$ possible demand vectors: $D_1, D_2,\dots, D_{\lfloor N/s\rfloor}$ in the same law before. Keep the same expression of $\tilde X $, ${{\tilde X}_{\backslash p}} $ and $\tilde Z$ as the first case. We define $\tilde W = \{ {W_1},...,W_{(p-1)s},W_{ps+1}...,{W_{\left\lfloor {N/s} \right\rfloor s}}\}$ this time.
	
In this case, the constraint \eqref{eq19} is still true. Due to $s\leq l$, according to \eqref{eq5}, the constraint \eqref{eq20} should be rewritten as 
\begin{align}
	I(\tilde W;{X_p},\tilde Z) = 0,\label{eq23}
\end{align}
where $p \in \{ 1,2,...,\left\lfloor {N/s} \right\rfloor \} $. Therefore we now give derivation below:
\begin{align}
	(\lfloor {N/s}\rfloor s-s)F =& H(\tilde W) \notag\\
	=& I(\tilde W;\tilde X,\tilde Z) + H(\tilde W|\tilde X,\tilde Z)\notag\\
	\overset{(a)}{=}& I(\tilde W;\tilde X,\tilde Z) \notag\\
	=& I(\tilde W;{X_p},\tilde Z) + I(\tilde W;{{\tilde X}_{\backslash p}}|{X_p},\tilde Z)\notag\\
	\overset{(b)}{=}& I(\tilde W;{{\tilde X}_{\backslash p}}|{X_p},\tilde Z)\notag\\
	\le& H({{\tilde X}_{\backslash p}})\notag\\
	\le& \sum\limits_{j=1,j\not=p}^{\lfloor {N/s}\rfloor}H(X_{D_j})\notag\\
	=&(\left\lfloor {N/s} \right\rfloor  - 1)RF.\label{eq24}
	\end{align}
	Here $(a),(b)$ are due to the constraints in \eqref{eq19},\eqref{eq23}. Therefore, from \eqref{eq24}, we can get a series of outer bounds,
	\begin{align}
	R \ge s ,
\end{align}
where $s\in\{1,2,\dots,\min(\lfloor {N/2} \rfloor,K)\}$. The lower bound of the optimal achievable communication load rate ${R^*}(M)$ could be concluded as
\begin{align}
	R^*(M)\!\ge \!R_s^*(M)&=\max \limits_{s\in\{1,2,\dots,\min(\lfloor N/2 \rfloor,K)\}}s\notag\\
	&=\min(\lfloor N/2 \rfloor,K).\label{eq27}
\end{align}
Combining \eqref{eq26} and \eqref{eq27}, the Theorem \ref{T2} is proved.

\section{Proof of Corollary \ref{C1}}\label{BC}
In this subsection, we prove the Corollary \ref{C1}. We compare lower bound of $R^*(M)$ in Theorem \ref{T2} to the rate $R^C(M)$ achieved by the proposed scheme in Theorem \ref{T1} in three cases: $l\geq \min(\lfloor N/2 \rfloor,K)$, $\frac{K+1}{12}\leq l< \min(\lfloor N/2 \rfloor,K)$ and $l \leq \frac{K+1}{12}$.

Assume that $l\geq \min(\lfloor N/2 \rfloor,K)$. We have
\begin{align}
    R^C(M)& \leq R^C(1)=K,\notag\\
	R^*(M)& \ge R_s^*(M)=\min(\lfloor N/2 \rfloor,K), \notag
\end{align}
by Theorem \ref{T1} and Theorem \ref{T2}. Combining these yields
\begin{align}
    \frac{R^C(M)}{R^*(M)}\leq\frac{R^C(1)}{R_s^*(M)}=\frac{K}{\min(\lfloor N/2 \rfloor,K)}\leq 12,\label{REGION1}
\end{align}
for $l\geq \min(\lfloor N/2 \rfloor,K)$ and $\frac{K}{N}\leq 3$.

Assume next that $\frac{K+1}{12}\leq l< \min(\lfloor N/2 \rfloor,K)$. Setting $s=l$, we have
\begin{align}
	R^*(M)& \ge R_s^*(M)\ge \frac{l\lfloor N/l\rfloor-l-(l-l)M}{\lfloor N/l \rfloor-1} =l, \notag
\end{align}
by Theorem \ref{T2}. Combining these yields
\begin{align}
    \frac{R^C(M)}{R^*(M)}\leq\frac{R^C(1)}{l}\leq\frac{K}{\frac{K+1}{12}}\leq 12,\label{REGION2}
\end{align}
for $\frac{K+1}{12}\leq l< \min(\lfloor N/2 \rfloor,K)$ and $\frac{K}{N}\leq 3$.

Keep to assume that $l \leq \frac{K+1}{12}$. We view the achievable rate $R$ in Theorem \ref{T1} as a function of $t$, 
\begin{align*}
    R^C(t)&=\frac{\binom{K}{t+1}}{\binom{K-l}{t}}=\frac{K!(K-l-t)!}{(t+1)(K-l)!(K-t-1)!}\\
          &=\frac{K-t}{t+1}\cdot\prod\limits_{j = 0}^{t-1} {\frac{K-j}{K-l-j}},
\end{align*}
for $t\in \{0,1,\dots,\max(\left\lceil {\frac{{K + 1}}{l} - 2} \right\rceil,0)\}$. The rate $R^C(t)$ decreases as $t$ increases (proof omitted for space constraints). Thus, we have 
\begin{align*}
    R^C(t)&\leq R^C(\lfloor \frac{K+1}{10l} \rfloor)\\
          &=(\frac{K-t}{t+1}\cdot\prod\limits_{j = 0}^{t-1} {\frac{K-j}{K-l-j}})_{t=\lfloor \frac{K+1}{10l} \rfloor}\\
          &\leq(\frac{K-t}{t+1}\cdot[\frac{K(K-t+1)}{(K-l)(K-l-t+1)}]^{\frac{t}{2}})_{t=\lfloor \frac{K+1}{10l} \rfloor}\\
          &=(\frac{K-t}{t+1}\cdot[\frac{K}{K-l}]^{\frac{t}{2}}\cdot[\frac{K-t+1}{K-l-t+1}]^{\frac{t}{2}})_{t=\lfloor \frac{K+1}{10l} \rfloor}\\
          &\leq\frac{K-(\frac{K+1}{10l}-1)}{(\frac{K+1}{10l}-1)+1}\cdot
           (\frac{\frac{K}{l} }{\frac{K}{l} -1})^{\frac{1}{20l}}\cdot
           (\frac{\frac{K}{l} }{\frac{K}{l} -1})^{\frac{K}{l} \cdot \frac{1}{20} }\\
          &\qquad\qquad \cdot [\frac{(1-\frac{1}{10l})\cdot\frac{K+1}{l} }{(1-\frac{1}{10l})\cdot\frac{K+1}{l}-1}]^{\frac{K+1}{l} \cdot \frac{1}{20} }\\
          &\leq (10l-1)\cdot
           (\frac{11}{11-1})^{\frac{1}{20}}\cdot
           (\frac{11 }{11 -1})^{11 \cdot \frac{1}{20} }\\
          &\qquad\qquad \cdot [\frac{(1-\frac{1}{10})\cdot 12 }{(1-\frac{1}{10})\cdot 12-1}]^{12 \cdot \frac{1}{20} }\\
          &= 1.12778(10l-1), 
\end{align*}
for $\lfloor \frac{K+1}{10l} \rfloor \leq t\in \{0,1,\dots,\max(\left\lceil {\frac{{K + 1}}{l} - 2} \right\rceil,0)\}$ when $l \leq \frac{K+1}{12}$. Combining these yields
\begin{align}
    \frac{R^C(M)}{R^*(M)}\leq\frac{1.12778(10l-1)}{l}\leq 12,\label{REGION3}
\end{align}
for $l \leq \frac{K+1}{12}$ and $t\geq\lfloor \frac{K+1}{10l} \rfloor$. 

Combining \eqref{REGION1},\eqref{REGION2} and \eqref{REGION3} yields
\begin{align*}
    \frac{R^C(M)}{R^*(M)}\leq 12,
\end{align*}
for every achievable rate pair $(M,R^C(M))$ in Theorem \ref{T1} when $t\geq\lfloor \frac{K+1}{10l} \rfloor$ and $\frac{K}{N}\leq 3$. The Corollary \ref{C1} is proved. It shows that more than $9/10$ points in Theorem \ref{T1} can maintain multiplicative gap within the factor $12$. It means the inner bound in Theorem \ref{T1} is order optimal in general.

\section{Proof of Corollary \ref{C2}}
Corollary \ref{C2} can be proved easily. For every $n\in[N]$, 	
\begin{footnotesize}
	\[\left[\! {\begin{array}{*{20}{c}}
	{\tilde{W}_{n,\mathcal{T}_{1}}}\\
	{\tilde{W}_{n,\mathcal{T}_{2}}}\\
	\vdots \\
	{\tilde{W}_{n,\mathcal{T}_{G}}}
	\end{array}}\! \right]{\!\rm{ = }\!}\left[\! {\begin{array}{*{20}{c}}
	{c_{1,1}}&{c_{1,2}}&\cdots &{c_{1,G}} \\
	{c_{2,1}}&{c_{2,2}}&\cdots &{c_{2,G}} \\
	\vdots     &\vdots     & \ddots&     \vdots \\
	{c_{G,1}}&{c_{G,2}}&\cdots &{c_{G,G}}  \\
	\end{array}}\! \right]\!\left[ \!{\begin{array}{*{20}{c}}
	{W_{n,1}}\\
	\vdots \\
	{W_{n,P}}\\
	{Y_{n,1}}\\
	\vdots \\
	{Y_{n,Q}}
	\end{array}}\! \right]{\!\rm{ = \!C}\!}\left[ {\begin{array}{*{20}{c}}
	{W_{n,1}}\\
	\vdots \\
	{W_{n,P}}\\
	{Y_{n,1}}\\
	\vdots \\
	{Y_{n,Q}}
	\end{array}}\! \right],\]
\end{footnotesize}
where $C$ is a Cauchy matrix in Galois field. Its element $c_{i,j}=\frac{1}{x_i+y_j}$, where both $x_i$ and $y_j$ are the distinct elements of Galois field GF($2^m$) that satisfies $2^m \geq 2G$ for every $i,j \leq G$. 

On the one hand, the Cauchy matrix is always full rank and the inverse matrix $C^{-1}$ is available. Thus, if obtaining all the $\tilde{W}_{n,\mathcal{T}}$, we can multiply them by $C^{-1}$ to decode all the $W_{n,p}$ and $Y_{n,q}$. Equation \eqref{C2eq1} is true. 

On the other hand, we can pick up any $Q$ of $G$ blocks and form the collection $\{\tilde{W}_{n,\mathcal{T}_{i_{1}}},\tilde{W}_{n,\mathcal{T}_{i_{2}}},...,\tilde{W}_{n,\mathcal{T}_{i_{Q}}}\}$, where $\{i_1,i_2,\dots,i_Q\}$ is one subset of $[G]$. So
\begin{footnotesize}
	\[\left[ {\begin{array}{*{20}{c}}
		{\tilde{W}_{n,\mathcal{T}_{i_{1}}}}\\
		{\tilde{W}_{n,\mathcal{T}_{i_{2}}}}\\
		\vdots \\
		{\tilde{W}_{n,\mathcal{T}_{i_{Q}}}}
		\end{array}} \right]
		{\rm{ = }}\left[ {\begin{array}{*{20}{c}}
		{c_{i_1,1}}&{c_{i_1,2}}&\cdots &{c_{i_1,G}} \\
		{c_{i_2,1}}&{c_{i_2,2}}&\cdots &{c_{i_2,G}} \\
		\vdots     &\vdots     & \ddots&     \vdots \\
		{c_{i_Q,1}}&{c_{i_Q,2}}&\cdots &{c_{i_Q,G}}  \\
		\end{array}} \right]\left[ {\begin{array}{*{20}{c}}
		{W_{n,1}}\\
		\vdots \\
		{W_{n,P}}\\
		{Y_{n,1}}\\
		\vdots \\
		{Y_{n,Q}}
		\end{array}} \right]\]\\
		\[
		{\rm{ = }}\left[ V_1 V_2 \right]\left[ {\begin{array}{*{20}{c}}
		{W_{n,1}}\\
		\vdots \\
		{W_{n,P}}\\
		{Y_{n,1}}\\
		\vdots \\
		{Y_{n,Q}}
		\end{array}} \right]{\rm{ = }}V_1\left[ {\begin{array}{*{20}{c}}
		{W_{n,1}}\\
		\vdots \\
		{W_{n,P}}
		\end{array}} \right]+V_2\left[ {\begin{array}{*{20}{c}}
		{Y_{n,1}}\\
		\vdots \\
		{Y_{n,Q}}
		\end{array}} \right].\]		
\end{footnotesize}

If the collection could leak the information, there should be one non-zero vector $H$ satisfies
\begin{align*}
    HV_1\not=O,HV_2=O,
\end{align*}
where $O$ is the zero vector. It means the interference from encryption key $Y$ could be eliminated. However, we know any submatrix of Cauchy matrix is still full rank. The rows of $V_2$ are linearly independent, which implies the non-existence of such $H$. Thus, the collection is impossible to leak any information. Equation \eqref{C2eq2} is true. The Corollary \ref{C2} is proved.

\end{document}